%% file: manuscript.tex
\documentclass[twocolumn]{aastex631}

\usepackage{multirow}
\usepackage{amsmath}

\begin{document}

\title{AN INTERMEDIATE MASS BLACK HOLE HIDDEN BEHIND THICK OBSCURATION}

\author[0000-0001-9379-4716]{Peter~G.~Boorman}
\affiliation{Cahill Center for Astrophysics, California Institute of Technology, 1216 East California Boulevard, Pasadena, CA 91125, USA}

\author[0000-0003-2686-9241]{Daniel~Stern}
\affiliation{Jet Propulsion Laboratory, California Institute of Technology, Pasadena, CA 91109, USA}

\author[0000-0002-9508-3667]{Roberto~J.~Assef}
\affiliation{Instituto de Estudios Astrof\'{i}sicos, Facultad de Ingenier\'{i}a y Ciencias, Universidad Diego Portales, Av. Ej\'{e}rcito Libertador 441, Santiago, Chile}

\author[0000-0002-9807-4520]{Abhijeet~Borkar}
\affiliation{Astronomical Institute, Academy of Sciences, Bo\v{c}n\'{i} II 1401, CZ-14131 Prague, Czech Republic}

\author[0000-0002-8147-2602]{Murray~Brightman}
\affiliation{Cahill Center for Astrophysics, California Institute of Technology, 1216 East California Boulevard, Pasadena, CA 91125, USA}

\author[0000-0003-0426-6634]{Johannes~Buchner}
\affiliation{Max Planck Institute for Extraterrestrial Physics, Giessenbachstrasse, 85741 Garching, Germany}
\affiliation{Excellence Cluster Universe, Boltzmannstr. 2, D-85748, Garching, Germany}

\author[0000-0002-4945-5079]{Chien-Ting~Chen}
\affiliation{Science and Technology Institute, Universities Space Research Association, Huntsville, AL 35805, USA}
\affiliation{Astrophysics Office, NASA Marshall Space Flight Center, ST12, Huntsville, AL 35812, USA}

\author[0000-0001-5857-5622]{Hannah~P.~Earnshaw}
\affiliation{Cahill Center for Astrophysics, California Institute of Technology, 1216 East California Boulevard, Pasadena, CA 91125, USA}

\author[0000-0002-4226-8959]{Fiona~A.~Harrison}
\affiliation{Cahill Center for Astrophysics, California Institute of Technology, 1216 East California Boulevard, Pasadena, CA 91125, USA}

\author[0000-0003-1994-5322]{Gabriele~A.~Matzeu}
\affiliation{Quasar Science Resources SL for ESA, European Space Astronomy Centre (ESAC), Science Operations Department, 28692, Villanueva de la Ca\~{n}ada, Madrid, Spain}

\author[0000-0001-8640-8522]{Ryan~W.~Pfeifle}
\altaffiliation{NASA Postdoctoral Program Fellow}
\affiliation{X-ray Astrophysics Laboratory, NASA Goddard Space Flight Center, Greenbelt, MD 20771, USA}
\affiliation{Oak Ridge Associated Universities, NASA NPP Program, Oak Ridge, TN 37831, USA}

\author[0000-0001-5231-2645]{Claudio~Ricci}
\affiliation{Instituto de Estudios Astrof\'{i}sicos, Facultad de Ingenier\'{i}a y Ciencias, Universidad Diego Portales, Av. Ej\'{e}rcito Libertador 441, Santiago, Chile}
\affiliation{Kavli Institute for Astronomy and Astrophysics, Peking University, Beijing 100871, China}

\author[0000-0003-2931-0742]{Ji\v{r}\'{i} Svoboda}
\affiliation{Astronomical Institute, Academy of Sciences, Bo\v{c}n\'{i} II 1401, CZ-14131 Prague, Czech Republic}

\author[0000-0003-3638-8943]{N\'{u}ria Torres-Alb\`{a}}
\altaffiliation{GECO Fellow}
\affiliation{Department of Physics and Astronomy, Clemson University, Kinard Lab of Physics, Clemson, SC 29634, USA}
\affiliation{Department of Astronomy, University of Virginia, P.O. Box 400325, Charlottesville, VA 22904, USA}

\author[0000-0002-5208-1426]{Ingyin Zaw}
\affiliation{Center for Astro, Particle, and Planetary Physics (CAP3), New York University Abu Dhabi, P.O. Box 129188, Abu Dhabi, UAE}



\begin{abstract}
Recent models suggest approximately half of all accreting supermassive black holes (SMBHs; $M_{\rm BH}$\,$\gtrsim$\,10$^{5}$\,M$_{\odot}$) are expected to undergo intense growth phases behind Compton-thick ($N_{\rm H}$\,$>$\,1.5\,$\times$\,10$^{24}$\,cm$^{-2}$) veils of obscuring gas. However, despite being a viable source for the seeding of SMBHs, there are currently no examples known of a Compton-thick accreting intermediate mass black hole (IMBH; $M_{\rm BH}$\,$\sim$\,10$^{2}$\,--\,10$^{5}$\,M$_{\odot}$). We present a detailed X-ray spectral analysis of IC\,750 -- the only AGN to-date with a precise megamaser-based intermediate mass $<$\,10$^{5}$\,M$_{\odot}$. We find the equivalent width of neutral 6.4\,keV Fe\,K$\alpha$ to be 1.9$^{+2.2}_{-1.0}$\,keV via phenomenological modelling of the co-added 177\,ks \textit{Chandra} spectrum. Such large equivalent widths are seldom produced by processes other than fluorescence from dense obscuration. We fit three physically-motivated X-ray spectral models to infer a range of possible intrinsic 2\,--\,10\,keV luminosity posteriors that encompass the systematic uncertainties associated with a choice of model. Despite a wide range of predicted intrinsic 2\,--\,10\,keV luminosities between $\sim$\,10$^{41}$ and 10$^{43}$\,erg\,s$^{-1}$, all three models agree that IC\,750 has a Compton-thick line-of-sight column density to $>$\,99\% confidence. Compton-thick obscuration is well-documented to impinge substantial bias on the pursuit of SMBH AGN. Our results thus provide the first indication that Compton-thick obscuration should also be properly considered to uncover and understand the IMBH population in an unbiased manner.
\end{abstract}



\section{Introduction} \label{sec:intro}

It is well established that circum-nuclear obscuration and accretion rate are critical parameters that shape how supermassive black holes (SMBHs; $M_{\rm BH}$\,$\gtrsim$\,10$^{5}$\,M$_{\odot}$) grow. The majority of radiatively-efficient active galactic nuclei (AGN) are obscured, both in the local Universe (e.g., \citealt{Ricci17_bassV,Kammoun20,Boorman24a}) and out to higher redshifts (e.g., \citealt{Ueda14,Buchner15,Ananna19}). Furthermore, accretion rate appears to regulate circum-nuclear obscuration via feedback from radiation pressure on dusty gas \citep{Fabian08,Ishibashi18} that manifests as a prevalence of highly obscured sources with Eddington ratios $\lambda_{\rm Edd}$\,$\sim$\,0.01\,--\,0.1, and a deficit at higher Eddington ratios \citep{Ricci17edd,Ananna22,Ricci22}.

Intermediate mass black holes (IMBHs; $M_{\rm BH}$\,$\sim$\,10$^{2}$\,--\,10$^{5}$\,M$_{\odot}$; \citealt{Greene20}) provide a vital clue to understand how seed masses grew to $\gtrsim$\,10$^{7}$\,M$_{\odot}$ within $\lesssim$\,700 million years after the Big Bang, when the Universe was $\lesssim$\,5\% of its current age (e.g., \citealt{Banados18,Goulding23}). Even though massive black holes tend to be most detectable via observable signatures of remnant or ongoing accretion activity, today very few AGN with confirmed intermediate masses are known (e.g., \citealt{Greene07,Dong12,Reines13,Sartori15,Baldassare17,Reines22}). Of the already-small confirmed active IMBH population, very few have robust measurements of accretion rate nor sufficiently high line-of-sight column densities that are required to understand the role that obscuration and accretion rate play in the growth of IMBHs. However, uncovering more heavily obscured IMBH AGN is a difficult challenge. Any biases imposed on selecting accreting SMBHs in heavily obscured AGN (e.g., \citealt{Brandt15,Ricci15,Balokovic17,Hickox18,TorresAlba21}) would be exacerbated in the case of accreting IMBHs which are intrinsically lower luminosity and easily confused with competing star formation-related processes.

For example, current optical spectroscopic surveys searching for accreting IMBHs require any AGN emission to exceed the competing emission from star formation in the host galaxy (e.g., \citealt{Greene04,Moran14}). X-ray observations are less affected by host contamination and so can be a powerful tool for finding low-luminosity obscured AGN (e.g., \citealt{Mezcua16,Annuar17,Chen17b,Annuar20,Ansh23,Mohanadas23}). In particular \citet{Ansh23} present detailed broadband X-ray and multi-wavelength analysis of one of the highest obscuration dwarf AGN currently known, finding a substantial line-of-sight column density of $N_{\rm H}$\,=\,3.4\,--\,7.0\,$\times$\,10$^{23}$\,cm$^{-2}$. However to-date no intermediate mass AGN have been confirmed with a Compton-thick ($N_{\rm H}$\,$>$\,1.5\,$\times$\,10$^{24}$\,cm$^{-2}$) line-of-sight column density.\footnote{We define a Compton-thick threshold by the inverse of the Thomson optical depth, noting however that the threshold does depend on other factors such as metallicity; see \url{http://mytorus.com/mytorus-instructions.html} for more information.}

An additional challenge arises from attaining robust black hole mass measurements. Precise and accurate mass measurements for any type of obscured massive black hole are notoriously difficult (e.g., \citealt{Koss22}). One of the most accurate mass measurement techniques known arises from the Keplerian rotation of water vapour emission at 22\,GHz in megamasers (\citealt{Churchwell77}). Due to the requirement of nearly edge-on dense obscuration for detection, all known examples are obscured in X-rays with a substantial fraction found to be Compton-thick (e.g., \citealt{Greenhill08,Masini16,Brightman16,Brightman17}). Additional complexities associated with the X-ray spectral fitting of obscured AGN can be simplified with megamasers since the line-of-sight inclination can be frozen to edge-on, removing a large source of uncertainty in physically-motivated obscuration models (e.g., \citealt{Boorman24a}). Despite recent advancements related to megamaser emission detections in AGN (e.g., \citealt{Panessa20}), the number of sources with detected disk megamasers that are required for robust black hole mass measurements are small and almost always associated with black hole masses in the range $M_{\rm BH}$\,$\sim$\,$10^{6}$\,--\,$10^{8}$\,M$_{\odot}$ \citep{Masini16}.

Here we present the X-ray spectral analysis of IC\,750 (redshift-independent distance of 14.1\,Mpc; \citealt{Zaw20}) -- the first and currently only disk megamaser confirmed to host an IMBH. The target was identified by \citet{Chen17b} from the 40-month \textit{NuSTAR} serendipitous survey \citep{Lansbury17}, in which the source was found to have 22\,$\pm$\,11\,counts in the hard X-ray 8\,--\,24\,keV passband. Though the serendipitous detection was too faint with \textit{NuSTAR} for informed constraints from spectral fitting, the authors reported a line-of-sight column density in excess of $\sim$\,10$^{23}$\,cm$^{-2}$ derived from phenomenological fitting of the archival 14\,ks \textit{Chandra} observation. By fitting Keplerian rotation curves to the 22\,GHz megamaser data, \citet{Zaw20} then constrained the central black hole mass posterior to be enclosed within 4\,--\,14\,$\times$\,10$^{4}$\,M$_{\odot}$ with a mode of 7\,$\times$\,10$^{4}$\,M$_{\odot}$. The considerably low black hole mass for the IC\,750 deviates $\sim$\,2\,dex below standard scaling relations between black hole mass and stellar velocity dispersion and $\sim$\,1\,dex below relations with bulge mass and stellar mass (e.g., \citealt{Kormendy13,Greene20}). Most recently, \citet{Boorman24a} reported the co-added \textit{Chandra} spectrum of IC\,750, totalling 177\,ks of effective exposure time. The updated spectral quality of the co-added X-ray spectrum indicated a large Iron K$\alpha$ line with an equivalent width in excess of 2\,keV when fit with a simple phenomenological prescription for the underlying continuum. Such extreme fluorescence lines are currently exclusively observed in heavily Compton-thick AGN (e.g., \citealt{Levenson02,Boorman18}), strongly indicating IC\,750 to be the first Compton-thick intermediate mass AGN.

The paper is organised as follows: In Section~\ref{sec:data}, we extract spectra for every epoch of \textit{Chandra} data available for IC\,750, before co-adding. We then report the results of fitting the co-added \textit{Chandra} spectrum with three physically-motivated obscuration models in Section~\ref{sec:discussion} before comparing our results with other confirmed megamasers.

\section{Analysis} \label{sec:data}

\subsection{Chandra Observations}\label{sec:chandra}

\begin{figure*}
\centering
\includegraphics[width=1.\textwidth]{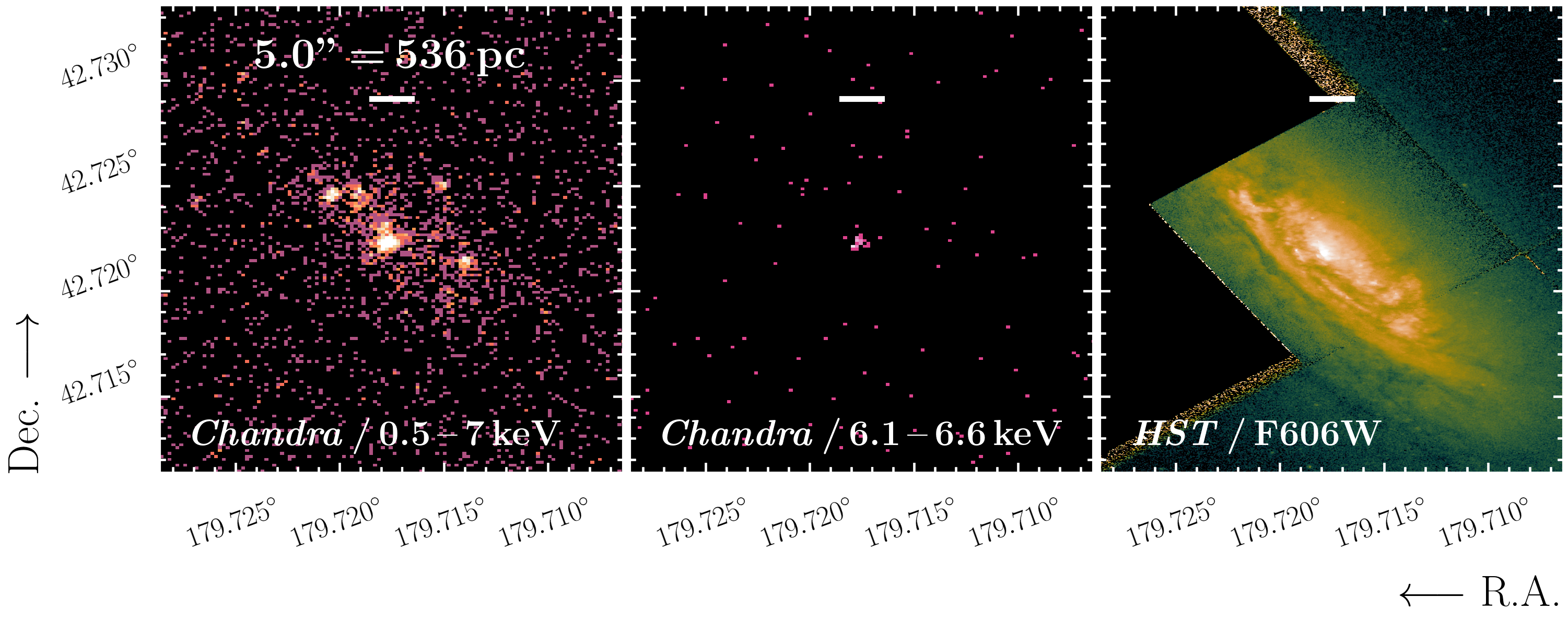}
\caption{\label{fig:images} \textit{Left:} Co-added 177\,ks exposure \textit{Chandra} image in the 0.5\,--\,7\,keV passband. \textit{Centre:} Co-added \textit{Chandra} image in the 6.1\,--\,6.6\,keV passband. The central AGN produces significant flux consistent with Fe\,K$\alpha$. \textit{Right:} Optical \textit{HST}/F606W image centred on the nucleus of IC\,750. All panels are matched in coordinates, and oriented such that north is up and east is to the left.
}
\end{figure*}

The central kiloparsec of the host galaxy of IC\,750 encompasses $\sim$\,10\,arcsec on the sky (see right panel of Figure~\ref{fig:images}). Since off-nuclear X-ray contaminants can have comparable observed flux to lower mass (and lower-luminosity) AGN (e.g., \citealt{Brightman18}), we focus solely on \textit{Chandra} data in this work due to its high angular resolution. Each \textit{Chandra} observation of IC\,750 was reprocessed using the Chandra Interactive Analysis of Observations (\textsc{ciao}\footnote{\url{https://cxc.cfa.harvard.edu/ciao}}; \citealt{Fruscione06}) command \texttt{chandra\_repro}.

We next used the \textsc{ciao} command \texttt{fluximage} to produce exposure-corrected images and exposure maps for each observation in the broad 0.5\,--\,7\,keV and narrow 6.1\,--\,6.6\,keV energy bands. The latter passband was chosen to encompass observed frame Fe\,K emission in the source. We used \texttt{reproject\_obs} to reproject all observations to a new reference point, before running \texttt{flux\_obs} to co-add the reprojected observations into individual exposure-corrected images. The resulting broadband and Fe\,K images are shown in the left and centre panels of Figure~\ref{fig:images}, respectively. Although a number of off-nuclear contaminants are revealed within $\sim$\,7\,--\,15\,arcsec of IC\,750 (see \citealt{Zaw20} for more information), the AGN is clearly spatially identified by its Fe\,K emission shown in the centre panel.

We extracted source\,$+$\,background counts for each observation separately from a circular region of 2\,arcsec radius centered on the source. The background counts were extracted from an annular region centered on the source with 36\,arcsec and 94\,arcsec inner and outer radii, respectively. The background regions were additionally chosen to exclude off-nuclear contaminants and any diffuse emission associated with the host galaxy. The source\,$+$\,background and background spectral files, as well as response and effective area files, were created using the \textsc{ciao} command \texttt{specextract}. The co-added \textit{Chandra} spectrum was then binned using the optimal prescription of \citet{Kaastra16} using \texttt{ftgrouppha}\footnote{\url{https://heasarc.gsfc.nasa.gov/lheasoft/help/ftgrouppha.html}}.

Table~\ref{tab:chan_info} shows that the count rates in the soft and hard bands are all consistent across the different epochs. We note that the decreasing soft band count rate over time is fully consistent with the effective area degradation of \textit{Chandra}. To search for any possible signs of spectral variability, we additionally fit each \textit{Chandra} spectrum with a phenomenological prescription (see Section~\ref{subsec:texrav} for more information). We then used quantile-quantile difference plots together with posterior predictive checks (see \citealt{Buchner23} for more information) to test how well the spectral fits to each epoch can explain the data of each other epoch. We find no significant evidence for spectral variability between epochs, so we co-added the spectra from all six observations using the \textsc{ftool} command \texttt{addspec}\footnote{\url{https://heasarc.gsfc.nasa.gov/ftools/caldb/help/addspec.txt}}. The final co-added spectrum that we use for all spectral fitting in this paper constitutes a net exposure of 177\,ks with a net count rate in the 0.5\,--\,8\,keV passband of 2.43\,$\pm$\,0.12 counts/ks corresponding to a total of 430 net counts (see Table~\ref{tab:chan_info} for the observed properties of the individual and co-added spectra).

\begin{table*}
\centering
\caption{\textit{Chandra} data used in this work.\label{tab:chan_info}}
\input{rate_info}\\
{\raggedright \textbf{Notes.} (1)--observation ID; (2)--Principle Investigator for the observation; (3)--observation start date and time; (4)--time difference relative to previous row; (5)--net exposure time in ks; (6), (7) and (8)--net count rate in counts per ks for the soft (0.5\,--\,2\,keV), hard (2\,--\,8\,keV) and broad (0.5\,--\,8\,keV) bands, respectively; (9), (10) and (11)--signal-to-noise in the soft, hard and broad bands, respectively, computed with the \texttt{gv\_significance} library of \citet{Vianello18}.\footnote{\url{https://github.com/giacomov/gv_significance}} The factor of $\sim$\,2 decrease in soft X-ray count rate of the later observations relative to the 2014 observation is due to the soft band degradation of \textit{Chandra}. The soft band fluxes are comparable across all epochs.
}
\end{table*}

\subsection{X-ray Spectral Fitting}
All X-ray spectral fitting presented in this paper was performed with \textsc{PyXspec} \citep{Gordon21}, the Python implementation of the X-ray Spectral Fitting software \textsc{Xspec} \citep{Arnaud96}. Spectral fits were carried out using the W-statistic\footnote{\url{https://heasarc.gsfc.nasa.gov/xanadu/xspec/manual/XSappendixStatistics.html}} (also known as the modified C-statistic; \citealt{Wachter79}). All parameter exploration was carried out with the Bayesian X-ray Analysis software package (\textsc{BXA} v4.0.5; \citealt{Buchner14,Buchner16}), which connects \textsc{PyXspec} to the nested sampling package \textsc{UltraNest} v4.0.5 \citep{Buchner21_ultranest}. To aid the computation time associated with sampling new points of higher likelihood throughout the fitting process, we use step samplers within \textsc{UltraNest} \citep{Buchner22}, initially starting from a given number of steps and doubling until convergence is achieved. As recommended by \citet{Buchner24}, we use the relative jump distance to verify our nested sampling chains have converged. Specifically, we quote the geometric mean random jump distance and the proportion of random jump distances above unity for each fit.

All spectral parameters are quoted as the maximum a posteriori value together with the 90\% highest density interval integrated from each corresponding marginalised posterior mode, unless stated otherwise. All luminosities reported for IC\,750 use the distance adopted by \citet{Zaw20} of 14.1\,Mpc, which was used in the calculation of the black hole mass. The distance was estimated by \citet{Zaw20} from the dynamics of the NGC\,4111 group that IC\,750 belongs to. For luminosities used from other works, we convert the luminosities to an equivalent distance of 14.1\,Mpc. The photon index of the intrinsic coronal emission for all spectral fits was assigned a Gaussian prior of mean 1.8 and standard deviation 0.15 and the cut-off energy was frozen to 300\,keV in all fits presented in this paper, in agreement with population studies of local AGN (e.g., \citealt{Balokovic17,Ricci17_bassV,Ricci18,Balokovic20}). All remaining parameters were assigned uniform or log-uniform priors depending on the nature of the parameter (i.e. whether it ranges over many orders of magnitude), unless stated otherwise.

\section{Results \& Discussion} \label{sec:discussion}

\begin{table*}
\centering
\caption{Parameters derived from X-ray spectral fitting of IC\,750.\label{tab:param_info}}
\input{params}\\
{\raggedright \textbf{Notes.}\\
Columns: (1)--physical property; (2)--parameter values derived from the phenomenological \texttt{texrav} model (note this model is least favoured -- see Section~\ref{subsec:texrav}); (3), (4) and (5)--parameter values derived for the \texttt{borus02}, \texttt{UXCLUMPY} and \texttt{warpeddisk} models, respectively; (6)--parameter units.\\
Rows: $^{a}$Temperature of the \texttt{APEC} component. $^{b}$Intrinsic luminosity of the \texttt{APEC} component in 0.5\,--\,2\,keV. $^{c}$Photon index. $^{d}$Line-of-sight column density. $^{e}$Thomson-scattered fraction. $^{f}$Observed 2\,--\,10\,keV flux. $^{g}$Observed 2\,--\,10\,keV luminosity. $^{h}$Intrinsic 2\,--\,10\,keV luminosity of the powerlaw. $^{i}$Ratio of intrinsic-to-observed 2\,--\,10\,keV flux. $^{j}$Bolometric luminosity derived with the bolometric correction of \citet{Brightman17}. $^{k}$Eddington ratio. $^{l}$ Number of steps used for the step sampler within \textsc{UltraNest}. $^{m}$The geometric mean of the random jump distance. $^{n}$The fraction of random jump distances above unity. $^{o}$Bayes Factor relative to the \texttt{texrav} model fit. We note that the positive values of all Bayes Factors indicates an improvement in the spectral fit with all models relative to \texttt{texrav}. Since the objective of this paper is not to constrain a geometry, we simply use this as a simplistic goodness-of-fit verification for our modelling.\\
}
\end{table*}

\subsection{Phenomenological Fitting}\label{subsec:texrav}
As shown in the centre panel of Figure~\ref{fig:images}, there is substantial emission in the 6.1\,--\,6.6\,keV energy range that is coincident with the nucleus of IC\,750. This is consistent with the findings of \citet{Boorman24a} who reported an Fe\,K$\alpha$ equivalent width in excess of 2\,keV derived from a simple powerlaw prescription for the underlying continuum.

To improve upon the simple prescription incorporated in that analysis, we first included the underlying reprocessed AGN continuum in a phenomenological manner. For solar abundances and typical obscurer covering factors, many X-ray radiative transfer simulations agree that the equivalent width of neutral Fe\,K$\alpha$ relative to the underlying Compton-scattered continuum is always $\geq$\,1\,keV when the line-of-sight column density exceeds the Compton-thick threshold \citep{Ikeda09,Murphy09,Brightman11,Tanimoto19}. However, the underlying continuum can often comprise additional components which may saturate the total detected line flux or give rise to artificially small equivalent widths in some cases (see discussion in \citealt{Levenson02,Gandhi17,Boorman18} for more information).

To aid the computation time involved with fitting, we generated a custom table of the \texttt{pexrav} model \citep[\texttt{texrav} hereafter]{Magdziarz95} to reproduce the Compton scattered continuum of IC\,750. The table model was created with an appropriate spectral resolution for our \textit{Chandra}/ACIS data, and assumes solar abundances with an edge-on inclination angle to match the megamaser in the source. Since we do not trace the peak of the Compton scattered continuum (i.e. the Compton hump at $\sim$\,20\,--\,30\,keV), we froze the relative normalisation of the reprocessed component to unity \citep{Balokovic17}.

\begin{figure}
\centering
\includegraphics[width=0.99\columnwidth]{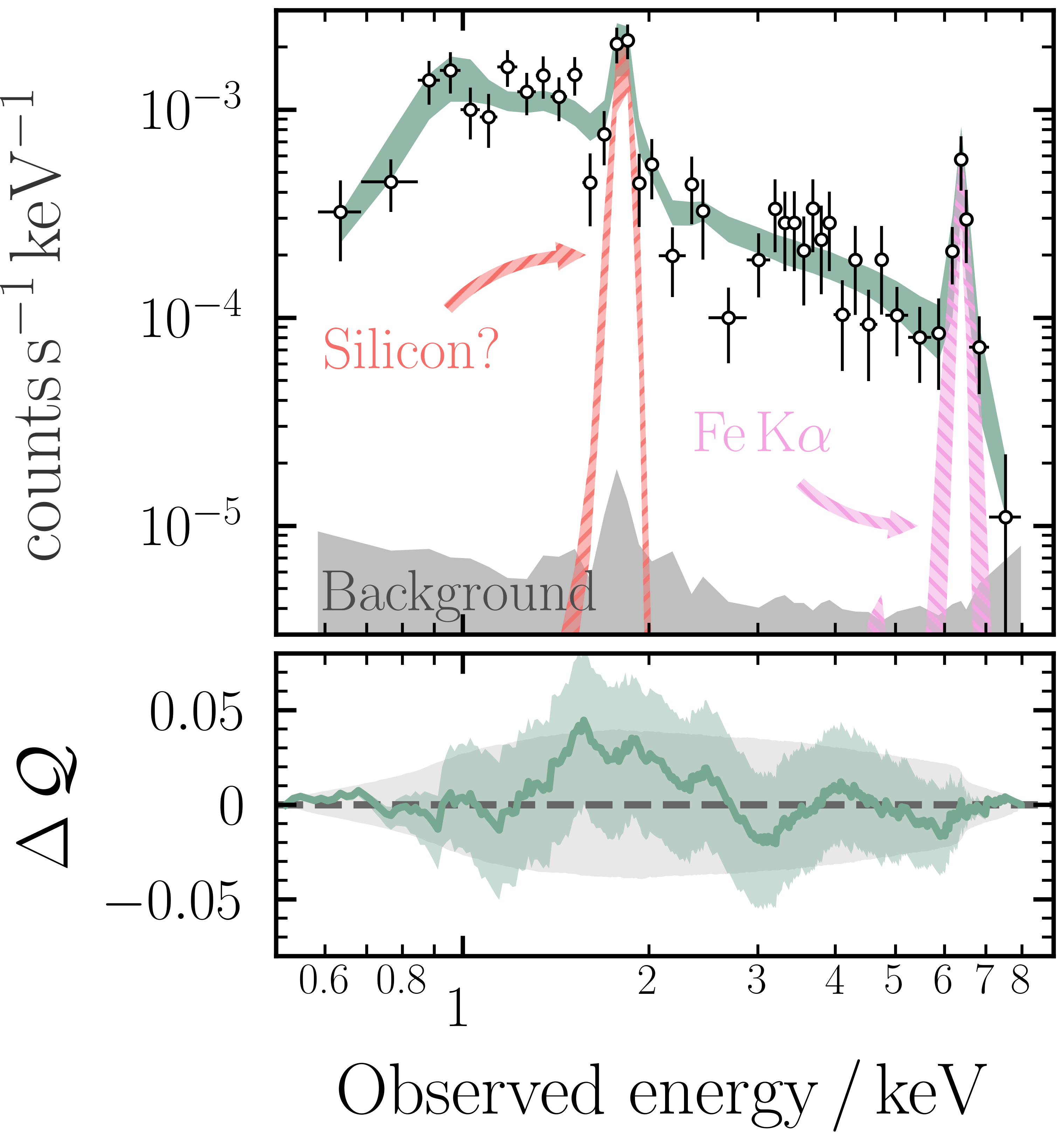}
\caption{\label{fig:texrav} \textit{(Top)} The co-added \textit{Chandra} data and corresponding \texttt{texrav}-based fit, folded with the instrument response. The background has been subtracted for visual clarity, but was not subtracted during fitting. The total posterior model range is shown in green, together with the posterior constraint on the Fe\,K$\alpha$ and tentative Silicon emission lines shown in pink and red hatched shaded regions, respectively. The instrument background is plotted with grey shading at the bottom of the panel. \textit{(Bottom)} Difference in the cumulative posterior model predicted counts and cumulative measured data counts are plotted as a function of energy with green shading. Simulated posterior predictive checks, showing the range expected if the derived posterior were true, is shown in grey. Since the green posterior is broadly bounded by the grey shaded region, we deem the fit as acceptable.
}
\end{figure}

Our phenomenological model setup was constructed to follow model B2 in Section 4.2.2 of \citet{Ricci17_bassV}. The primary X-ray coronal emission was modelled with a transmitted cut-off powerlaw. An additional cut-off powerlaw was included to reproduce a small fraction (forced to be $<$\,10\%) of intrinsic emission that escapes the circum-nuclear environment through a lower column density than the primary obscurer (e.g., \citealt{Gupta21}). A thermal \texttt{APEC} (Astrophysical Plasma Emission Code, v.12.10.1; \citealt{Smith01}) component was then used to explain any surplus of soft X-ray flux $\lesssim$\,3\,keV, as well as a Gaussian line to fit Iron emission between 6\,--\,7\,keV at the redshift of the source. On inspection, we noticed residuals around $\sim$\,1.8\,--\,1.9\,keV in the individual and co-added spectra, consistent with some highly-ionised species of Silicon emission (e.g., Si\,\textsc{xviii}). We thus included an additional Gaussian to provide a good overall fit to the entire spectrum. \citet{Liu19} showed that polar gas is expected to produce significant Si\,K$\alpha$ emission at 1.74\,keV that can be blue-shifted with respect to the host galaxy if in an outflowing configuration. However, since the line also appears strongly in the observed background (c.f. Figure~\ref{fig:texrav}), we adhere caution to inferring physical properties related to any tentative Silicon emission. Future high-spectral resolution observations (in particular when spatially-resolved) will enable a better understanding of low-energy outflow signatures in obscured AGN (see e.g., \citealt{Matzeu22,Gandhi22,Barret23}).

The corresponding fit with the \texttt{texrav}-based model in folded units is shown in the upper panel of Figure~\ref{fig:texrav} together with the spectral posterior ranges on the full model, Fe\,K$\alpha$ component and Silicon component. The bottom panel of Figure~\ref{fig:texrav} presents the posterior cumulative difference between normalised detected counts and normalised model-predicted counts (i.e. the quantile-quantile difference; see \citealt{Buchner23} for more information). The green thick line and associated 90\% confidence shading shows the cumulative difference of the model posterior derived from fitting the real data. We perform posterior predictive checks by simulating a large number of posterior rows with the same observational setup as the real co-added spectrum (i.e. background, response and effective area files, exposure time). The corresponding range in cumulative difference curves under the assumption that the posterior (with its associated uncertainty) were correct defines the background grey shaded region, which broadly encompasses the solid green real posterior curve. Our posterior predictive checks for the \texttt{texrav} model thus show that the model is able to reproduce the data adequately.

Figure~\ref{fig:texrav} shows that the model requires significant neutral Fe\,K$\alpha$ emission, finding the rest frame line energy to be 6.40\,$\pm$\,0.03\,keV and observed equivalent width to be 1.9$^{+2.2}_{-1.0}$\,keV. Such a large equivalent width is broadly consistent with the very local Compton-thick population (see \citealt{Boorman18}), though we note a large posterior uncertainty associated with the lack of underlying continuum constraints at $>$\,10\,keV. Interestingly, the fit also yields a line-of-sight column density posterior of log\,$N_{\rm H}$\,/\,cm$^{-2}$\,=\,24.16$^{+0.83}_{-0.23}$, which is Compton-thick but consistent with being below the Compton-thick threshold (see Table~\ref{tab:param_info} for details of other parameters in the fit).

However, the slab-based geometry assumed in \texttt{pexrav} is a poor representation of the obscuration surrounding AGN (see discussion in e.g., \citealt{Balokovic17,Paltani17,Buchner21,Boorman24a}). Furthermore, the line-of-sight column density derived is decoupled from the level of Compton scattering which can give rise to strong parameter degeneracies (e.g., between the intrinsic continuum slope and column density). Thus column densities inferred with our \texttt{texrav}-based model should be met with caution. To address this concern, we next turned to physically-motivated models with self-consistent prescriptions for photoelectric absorption, fluorescence, and Compton scattering.

\subsection{Physically-motivated Modelling}\label{subsec:physical}

Using different physically-motivated models\footnote{We refer to models in which photoelectric absortion, fluorescence and Compton-scattering are self-consistently modelled as physically-motivated throughout the manuscript. We note however that some models (e.g., \texttt{borus02} and \texttt{UXCLUMPY}) do not have a known physical mechanism to create and maintain their geometries, and should thus be considered ad-hoc geometries combined with X-ray radiative transfer.} on the same obscured AGN spectrum can give significantly different parameter inference (e.g., \citealt{Brightman15,Lamassa19,Saha22,Kallova24,Boorman24a}, Boorman et al., in prep.). For this reason, we test three models described by distinct geometries: (1) \texttt{borus02} \citep{Balokovic18} which we use in its coupled configuration to replicate a uniform density sphere with parameterised polar cutouts. (2) \texttt{UXCLUMPY} \citep{Buchner19}, a clumpy configuration of variable size gas clouds and an additional Compton-thick inner ring. (3) \texttt{warpeddisk} \citep{Buchner21}, in which the primary obscurer arises from a Compton-thick disk with variable warp strength. The latter model holds particular physical interest since IC\,750 has been found to possess a $\sim$\,0.2\,pc diameter warped maser disk \citep{Zaw20}. For results arising from the X-ray spectral fitting with \texttt{MYtorus} \citep{Murphy09}, see Zaw et al., in prep.

All fits fundamentally included the same components as in the phenomenological \texttt{texrav}-based fit described in Section~\ref{subsec:texrav}, although the \texttt{texrav} component was replaced with the reprocessed (Compton scattered and fluorescence) spectrum and the intrinsic powerlaw was replaced with the transmitted component of each model. The Thomson-scattered component used to reproduce the soft excess in obscured AGN was included, but for \texttt{UXCLUMPY} and \texttt{warpeddisk} this component is provided as its own table model replicating Thomson scattering from optically thin material.

\begin{figure*}
\centering
\includegraphics[width=0.99\textwidth]{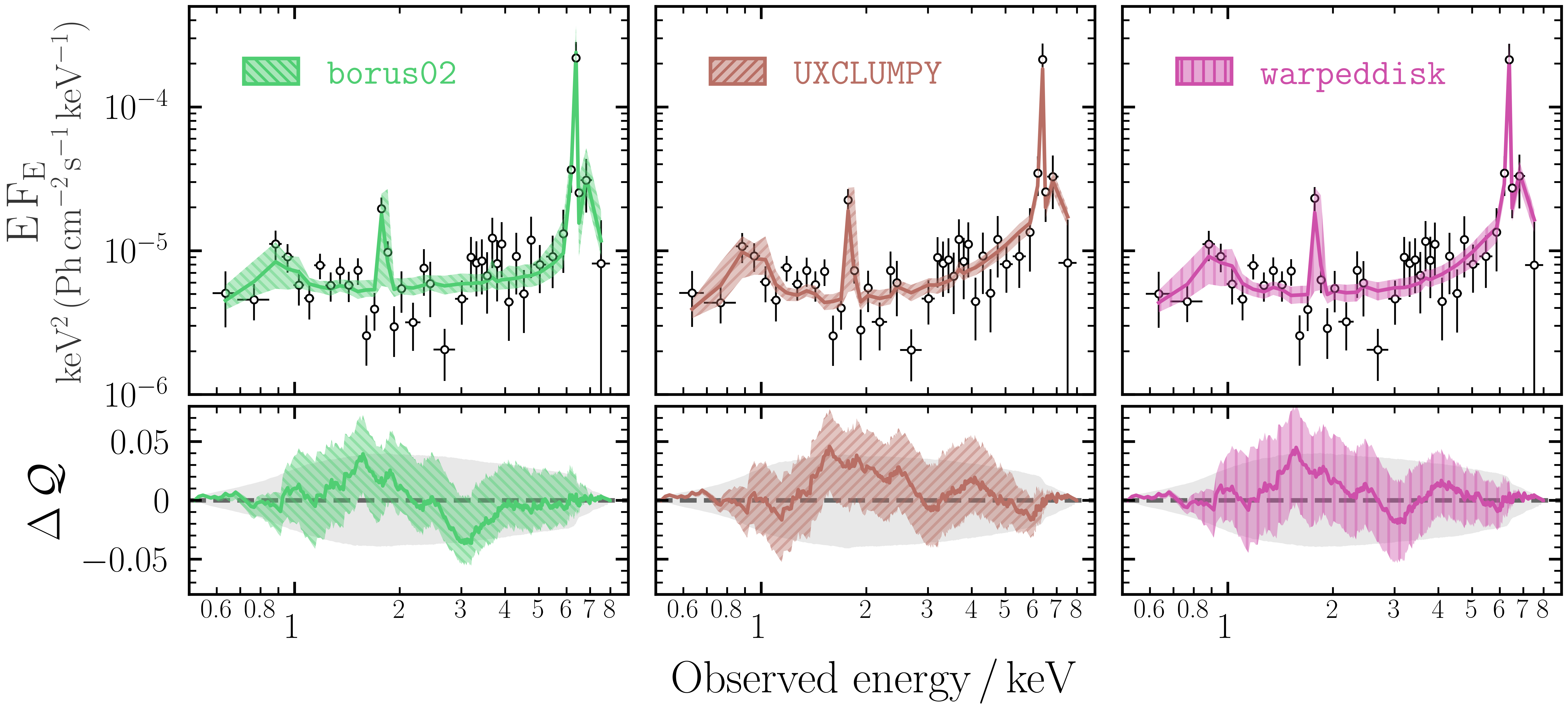}
\caption{\label{fig:eeuf} Unfolded spectral posteriors for each physically-motivated obscuration model fit to the co-added \textit{Chandra} data of IC\,750. Lower panels show the difference in cumulative (and normalised) model and data counts as a quantile-quantile difference plot. The grey oval-shaped regions present the 90\% confidence range expected by simulating the posterior model many times with the instrumental setup of the co-added data.
}
\end{figure*}

Figure~\ref{fig:eeuf} shows the unfolded spectral posteriors for each of the three physically-motivated models. The strong Fe\,K$\alpha$ line is reproduced well by each model with somewhat similar overall contributions to the full $\sim$\,0.5\,--\,8\,keV passband being fit over. The quantile-quantile difference panels for each physically-motivated model fit appear to show a similar shape. The characteristic `S' shape that is visible with an inflection at $\gtrsim$\,2\,keV is due to the apparent deficit of signal between $\sim$\,2\,--\,3\,keV. Such a flux deficit is somewhat reminiscent of an absorption feature, but our posterior predictive checks (grey shaded region in the lower panels of Figure~\ref{fig:eeuf}) show that such a feature is not significant given the uncertainty associated with the observed data. It is also important to note that the \textit{Chandra} data being fit is co-added from many different epochs, such that a flux deficit of the level shown could plausibly be a stochastic fluctuation. An interesting alternative interpretation could be that the flux deficit actually traces the underlying continuum with an additional broad emission feature between $\sim$\,3\,--\,4\,keV. Such emission features have received significant attention in the literature as a signature of sterile neutrino decay tracing dark matter (e.g., \citealt{Bulbul14}). However, all the caveats mentioned previously (i.e. posterior predictive checks and artefacts arising from co-adding) are just as pertinent for this alternative possibility as for the presence of absorption features. Future deeper observations with \textit{Chandra} could help discern any faint absorption or emission features in greater detail.

\begin{figure}
\centering
\includegraphics[width=0.99\columnwidth]{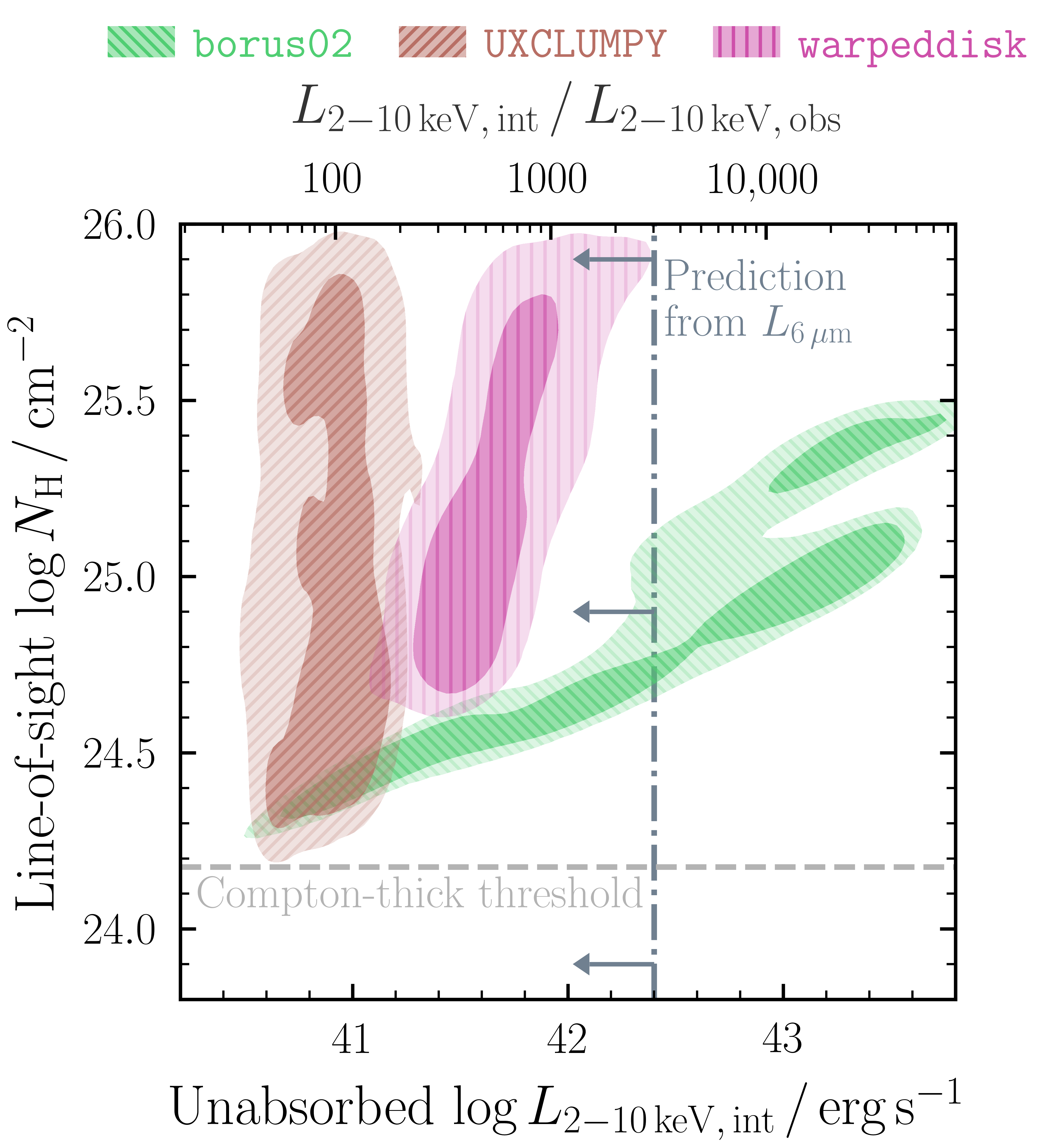}
\caption{\label{fig:LXvNH} Two dimensional posterior contours on intrinsic 2\,--\,10\,keV AGN luminosity vs. line-of-sight column density. The vertical dot-dashed line shows the upper limit on 2\,--\,10\,keV luminosity predicted from the mid-infrared 6\,$\mu$m luminosity, whereas the horizontal dashed line shows the Compton-thick threshold. Though the predicted intrinsic luminosities are very discrepant amongst different models, all converge on a Compton-thick line-of-sight column density to $>$\,99\% confidence.
}
\end{figure}

Figure~\ref{fig:LXvNH} shows the corresponding posterior fits in terms of the predicted intrinsic AGN luminosity in the extrapolated 2\,--\,10\,keV band vs. the predicted line-of-sight column density, both in logarithmic units. Whilst all three models unanimously predict line-of-sight column densities above the Compton-thick limit, each model reproduces a unique range in predicted intrinsic luminosities, the vast majority of which are inconsistent with one another. \texttt{borus02} has the widest range in allowable intrinsic AGN luminosities, extending up to $\sim$\,3\,$\times$\,10$^{43}$\,erg\,s$^{-1}$. However, given that the observed 2\,--\,10\,keV luminosity of IC\,750 is $\sim$\,8.3\,$\times$\,10$^{38}$\,erg\,s$^{-1}$, this would imply a significantly larger correction than the typical observed-to-intrinsic 2\,--\,10\,keV luminosity fractions found for Compton-thick AGN of $\sim$\,1.5\,--\,3 orders of magnitude (though this is also model dependent -- see e.g., \citealt{Boorman24a}).

\subsubsection{Isotropic Indicators of AGN Power}\label{subsec:istropic}
Given the relative discrepancies in predicted intrinsic luminosities of the three models being tested, it is necessary to investigate if any multi-wavelength estimators of intrinsic AGN power could potentially agree with a given model (e.g., \citealt{Gandhi14,Boorman16,Annuar17}).

The optical [O\,{\sc iii}] line flux is known to be correlated with intrinsic AGN power when the observed line is dominated by unobstructed emission originating from the narrow line region (e.g., \citealt{Panessa06,Berney15,Malkan17}). \citet{Zaw20} reports a de-reddened [O\,{\sc iii}] luminosity of log\,$L_{[{\rm O}\,\textsc{iii}]}$\,/\,erg\,s$^{-1}$\,$\sim$\,38.9, which is equivalent to a predicted intrinsic X-ray luminosity of log\,$L_{2-10\,{\rm keV},\,[{\rm O}\,\textsc{iii}]}$\,/\,erg\,s$^{-1}$\,$\sim$\,40.3 when using the best-fit to the optically-selected Seyfert~2 sample of \citet{Panessa06}. Measuring the true AGN-powered [O\,{\sc iii}] line flux is known to be uncertain due to a number of reasons including observing conditions, instrument configuration, spectral extraction method, stellar contaminating processes in the host galaxy and even the selection method by which a given sample is targeted (see discussion in \citealt{Goulding09,Ueda15,Berney15,Greenwell21,Greenwell22,Greenwell24}). Also, relations between the [O\,{\sc iii}] and X-ray luminosities are often calibrated for higher-mass AGN with higher metallicities that may give rise to additional scatter when inferring luminosities for lower-mass systems. All three physically-motivated models predict intrinsic X-ray luminosities that are significantly greater than the [O\,{\sc iii}]-predicted intrinsic X-ray luminosity. One possibility is that the [O\,{\sc iii}] flux traces an earlier epoch of lower AGN activity that is causally disconnected from the more recent circum-nuclear X-ray emission detected by \textit{Chandra}. Such ionised line emission echos are not unheard of in the local Universe (e.g., \citealt{Sartori16,EsparzaArredondo20,Saade22,Pfeifle23}).

The near-to-mid infrared can be a more precise indicator of intrinsic X-ray power if the AGN-powered infrared emission can be isolated reliably (e.g., \citealt{Horst08,Gandhi09,Asmus14,Mateos15,Stern15,Chen17a,Pfeifle22}). Since the \textit{Wide-field Infrared Survey Explorer (WISE)} $W1-W2$ colour of IC\,750 is $\sim$\,0.38\,mag or 0.22\,mag using the instrumental profile-fit photometry or elliptical aperture magnitudes, respectively, and the source appears significantly extended in the \textit{WISE} images, near-to-mid infrared fluxes from \textit{WISE} are expected to be dominated by host galaxy processes \citep{Stern12,Assef18,Asmus20}.

Instead, \citet{Chen17b} present a broadband spectral energy distribution decomposition for IC\,750, finding a 6\,$\mu$m luminosity of log\,$L_{6\,\mu{\rm m}}$\,/\,erg\,s$^{-1}$\,$\sim$\,42.3 for the AGN component. A detailed look at the spectral energy distribution modelling of IC\,750, following the same approach as \citet{Chen17b} with the models of \citet{Assef10}, shows that the system is likely affected by host-obscuration beyond the level covered by the \citet{Assef10} models. We find that re-fitting the broadband data of IC\,750 by either including additional host reddening or by removing the photometry blue-ward of $i$-band (which is most affected by the host obscuration) results in best-fits with substantially lower AGN contributions ($\gtrsim$\,0.5\,dex). To be conservative, we thus adopt the 6\,$\mu$m luminosity of \citet{Chen17b} as an upper limit. 

By using the 6\,$\mu$m-to-X-ray relation for the lower-luminosity subset of \citet{Chen17a}, we derive a 6\,$\mu$m-predicted intrinsic 2\,--\,10\,keV luminosity of log\,$L_{2-10\,{\rm keV},\,6\,\mu{\rm m}}$\,/\,erg\,s$^{-1}$\,$\lesssim$\,42.4 for IC\,750. The predicted 6\,$\mu$m luminosity is consistent with all X-ray spectral models, but excludes the high-luminosity tail of the \texttt{borus02} model fit that would correspond to an unlikely intrinsic-to-observed 2\,--\,10\,keV flux ratio of $\sim$\,4 orders of magnitude.

\subsection{Bolometric Luminosity and Eddington Ratio}\label{subsec:eddrat}

Recent studies have suggested a strong link between Eddington ratio and circum-nuclear environment column density arising from the effective Eddington limit on dusty gas. The effect fundamentally predicts a minimum line-of-sight column density that can be long-lived at a given Eddington ratio \citep{Fabian08,Ricci17edd,Ishibashi18}. Megamasers are ideal sources to study with regard to the effective Eddington limit on dusty gas due to their high prevalence of obscuration and precise black hole mass measurements. The remaining factor that can prove difficult on a case-by-case basis is a reliable estimate of the bolometric luminosity.

We rely on the X-ray bolometric correction to estimate bolometric luminosity, $\kappa_{\rm bol}$\,=\,$L_{\rm bol}$\,/\,$L_{2-10\,{\rm keV}}$. Whilst the bolometric correction has been extensively studied for unobscured AGN (e.g., \citealt{Marconi04,Vasudevan09,Jin12,Duras20}), it is less well-characterised for obscured AGN (e.g., \citealt{Vasudevan10,Lusso12}). \citet{Brightman17} constrained the X-ray bolometric correction for a sample of ten local Compton-thick AGN with robust 2\,--\,10\,keV intrinsic flux estimates and bolometric luminosities derived from infrared torus modelling and/or integrated spectral energy distribution modelling. The authors found an X-ray bolometric correction for the full sample of log\,$\kappa_{\rm bol}$\,=\,1.44\,$\pm$\,0.12 (with an intrinsic scatter of $\sim$\,0.2\,dex) and no significant offset relative to bolometric corrections for less-obscured AGN. However, since our intrinsic luminosity constraints are consistent with those of the \citet{Brightman17} sample, any bolometric correction dependence with luminosity (e.g., \citealt{Marconi04}) is expected to have a minor effect. Whilst some studies report black hole mass dependence for the X-ray bolometric correction (e.g., \citet{Duras20}), the number of AGN with precise black hole mass measurements $\lesssim$\,10$^{5}$ are exceedingly low. We do note however that extrapolating the best-fit black hole mass vs. bolometric correction relation of \citet{Duras20} gives a negligible mass dependence for IC\,750.

\begin{figure}
\centering
\includegraphics[width=0.99\columnwidth]{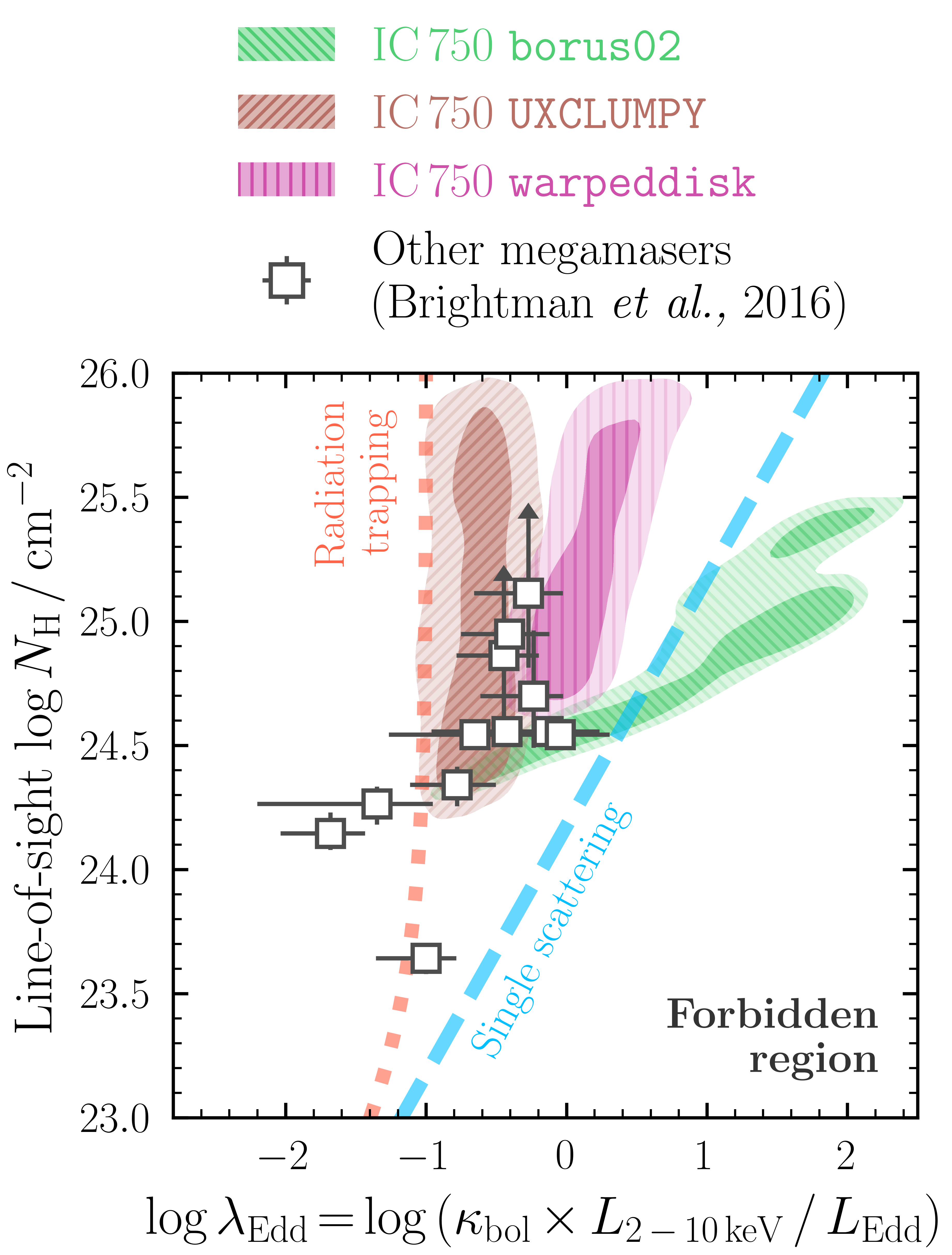}
\caption{\label{fig:effedd} The Eddington ratio posteriors derived for IC\,750 plotted against line-of-sight column density. The Eddington ratio posteriors were derived using the unabsorbed 2\,--\,10\,keV luminosities predicted from the physically-motivated obscuration modelling whilst propagating uncertainties in black hole mass and bolometric correction. The boundaries representing the effective Eddington limits on dusty gas in the regime of radiation trapping and the single scattering UV-thin regime from \citet{Ishibashi18} are plotted with dotted (red) and dashed (blue) lines, respectively. The sample of \textit{NuSTAR}-studied megamaser AGN from \citet{Brightman16} are plotted as dark grey squares with errorbars.
}
\end{figure}

Figure~\ref{fig:effedd} presents the line-of-sight column density vs. Eddington ratio derived for IC\,750 using the bolometric correction of \citet{Brightman17}, the black hole mass posterior from \citet{Zaw20} and the posteriors found for each X-ray spectral model. We additionally plot the megamaser sample of \citet{Brightman16} which includes Eddington ratio and line-of-sight column density estimations for 12 local Compton-thick AGN. Since \citet{Brightman16} used a bolometric correction of 10.0, we update the Eddington ratios from that sample to use the same bolometric correction that we use for IC\,750 from \citet{Brightman17}. Despite the sample of \citet{Brightman16} having black hole masses in the range $M_{\rm BH}$\,$\sim$\,10$^{6}$\,--\,10$^{8}$\,M$_{\odot}$ ($\sim$\,1\,--\,3\,dex higher than IC\,750), there is good agreement between the majority and the region of agreement for all three physically-motivated model posteriors with IC\,750. The theoretical boundary arising from the single-scattering-UV assumption of \citet{Fabian08} in which the infrared optical depth was assumed to be negligible is shown with a dashed blue line in Figure~\ref{fig:effedd} and restricts the high-power tail of the \texttt{borus02} posterior, akin to the 6\,$\mu$m-predicted intrinsic X-ray luminosity upper limit described in Section~\ref{subsec:istropic}.

The second red dotted threshold in Figure~\ref{fig:effedd} represents the effective Eddington limit when radiation trapping is considered in the infrared optically-thick regime \citep{Ishibashi18}, in which a given column density can be ejected by lower Eddington ratios, even in the Compton-thick regime, than allowed by the single-scattering-UV boundary. Interestingly the vast majority of sources from \citet{Brightman16}, as well as the posteriors for IC\,750, lie between the two thresholds. This could suggest that either (1) these Compton-thick AGN are actively blowing out their line-of-sight obscuration or (2) the Compton-thick megamaser AGN are in a more stable configuration in which the circum-nuclear obscuration is not ouflowing, and the radiation trapping threshold should lie at larger Eddington ratios for a given line-of-sight column density. \citet{Ishibashi18} provide a number of tertiary parameters that could shift the radiation trapping boundary to lie closer to the single scattering threshold, including the dust-to-gas ratio and the clumpiness of the obscurer that can decrease the overall fraction of radiation that is trapped. Additional uncertainty arises from the different X-ray models used by \citet{Brightman16} as compared to those fit to IC\,750 in this work, as well as the sheer scarceness of Compton-thick megamaser AGN in general. The ongoing effort to search for larger samples of megamasers (e.g., \citealt{Panessa20}) combined with next-generation broadband X-ray spectroscopic observations (from e.g., the \textit{High Energy X-ray Probe}, \textit{HEX-P}; \citealt{Madsen24,Boorman24a}) will enable a more complete understanding of the composition and structure of the circum-nuclear environment surrounding heavily obscured AGN across all mass scales.

\subsection{Circum-nuclear Obscuration Across the Mass Scale}\label{subsec:ctsamp}

\begin{figure*}
\centering
\includegraphics[width=0.99\textwidth]{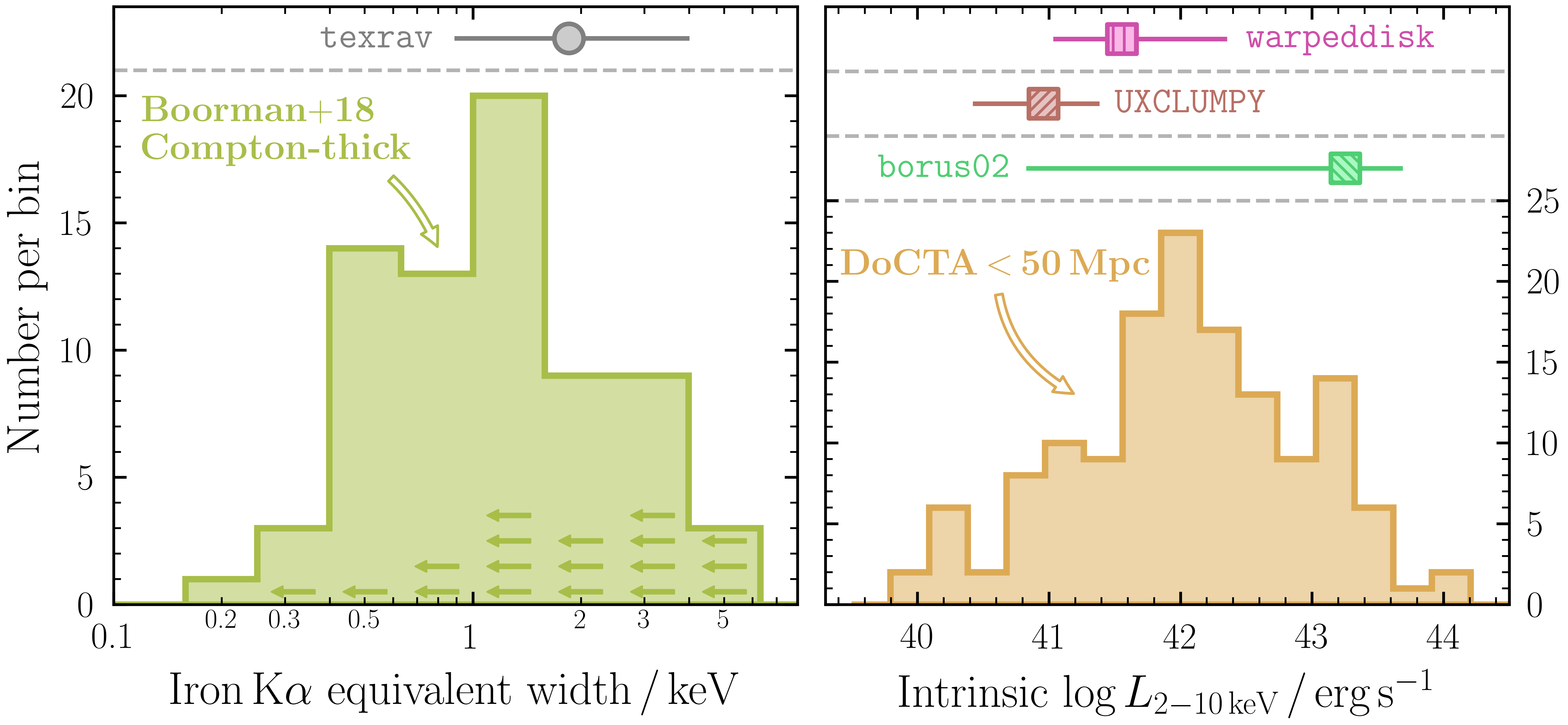}
\caption{\label{fig:ctcomp} \textit{(Left)} The distribution of Iron K$\alpha$ equivalent widths from the Compton-thick sample of \citet{Boorman18} with upper limits shown as solid green arrows. The large error bar at the top of the panel shows the \texttt{texrav}-based equivalent width measurement of Iron K$\alpha$ for IC\,750 from this work. \textit{(Right)} The distribution of 295 intrinsic luminosity measurements for 30 unique Compton-thick AGN in DoCTA within $D$\,$<$\,50\,Mpc \citep{Boorman24a}.
}
\end{figure*}

Bona fide intrinsically low-luminosity Compton-thick AGN with $L_{2-10\,{\rm keV}}$\,$\lesssim$\,10$^{41}$\,erg\,s$^{-1}$ are rare, but not unheard of in the local Universe (e.g., \citealt{Annuar17,Brightman18,daSilva21,Boorman24a}). A remaining fundamental question is thus whether IC\,750 is an outlier to the existing Compton-thick AGN population or if the properties of heavily obscured AGN can naturally extend into the intermediate mass regime. Figure~\ref{fig:ctcomp} shows the results for IC\,750 relative to two populations of Compton-thick AGN identified in the literature. The left panel shows the measured equivalent width of Iron K$\alpha$ for IC\,750 with the \texttt{texrav}-based model compared to the measured equivalent widths of the Compton-thick AGN population presented by \citet{Boorman18}. The right panel presents our intrinsic luminosity posterior measurements for IC\,750 compared to the Database of Compton-Thick AGN (DoCTA) presented by \citet{Boorman24a} within a volume of 50\,Mpc. Both the equivalent width derived with \texttt{texrav} and intrinsic luminosities covered by the three physically-motivated models are fully consistent with both populations, implying IC\,750 shares properties with the wider Compton-thick AGN population.

\section{Summary and Conclusions}\label{sec:summary}

We have studied the soft X-ray nuclear properties of IC\,750 -- the first and currently only IMBH with a megamaser-based mass measurement. Our main results are as follows:

\begin{enumerate}
    \item By co-adding all existing \textit{Chandra} observations for IC\,750, we find a number of soft X-ray point source contaminants within $\sim$\,7\,--\,15\,arcsec of the nucleus. However, by restricting our imaging to a narrow iron line range of 6.1\,--\,6.6\,keV, we show that the vast majority of emission is coincident with the nucleus (see Section~\ref{sec:chandra} and Figure~\ref{fig:images}). Future narrow band iron line imaging may thus prove to be a powerful technique for selecting more faint heavily obscured low mass AGN to infer properties of the underlying population. We stress that this is currently only possible with \textit{Chandra} owing to the requirement for $\lesssim$\,7\,arcsec resolution imaging over the $\sim$\,6\,--\,7\,keV passband. Future instruments with such capabilities include \textit{HEX-P} \citep{Madsen24,Pfeifle24}, \textit{AXIS} \citep{Reynolds23,Foord24}, \textit{Athena} \citep{Nandra13,Rau13,Meidinger20,Barret23} and \textit{Lynx} \citep{Gaskin19}.
    
    \item By co-adding spectra extracted from each individual epoch and fitting with a phenomenological model, we find a large equivalent width of neutral Fe\,K$\alpha$, constrained to be 1.9$^{+2.2}_{-1.0}$\,keV. We then fit three physically-motivated obscuration models which all find a Compton-thick line-of-sight column density to $>$\,99\% probability (see Section~\ref{sec:discussion} and Figures~\ref{fig:texrav} and~\ref{fig:LXvNH}). This is the first confirmation of a Compton-thick IMBH.
    
    \item The inferred intrinsic AGN power varies between the different physically-motivated models. Assuming the most up-to-date bolometric correction for Compton-thick AGN from \citet{Brightman17}, we show the source occupies Eddington ratios between $\sim$\,0.1\,--\,1, which exceed the effective Eddington limit for dusty gas when radiation trapping is considered \citep{Ishibashi18}. Similar results are found for other Compton-thick megamaser AGN in the literature, suggesting that the obscuration may be outflowing or the boundary should be updated for Compton-thick column densities (see Section~\ref{subsec:eddrat} and Figure~\ref{fig:effedd}). Approved observations with \textit{NuSTAR} \citep{Harrison13} will provide vital constraints on the circum-nuclear environment and intrinsic AGN power that are not attainable with spectroscopy solely at $<$\,8\,keV. However, currently only \textit{HEX-P} will be capable of providing well-matched simultaneous constraints over $\sim$\,0.3\,--\,80\,keV for similarly faint AGN in the future (e.g., \citealt{Boorman24a,Civano24}).
    
    \item We compare our results for IC\,750 to the Compton-thick Iron line equivalent width sample of \citet{Boorman18} as well as all confirmed Compton-thick AGN from DoCTA \citep{Boorman24a} within 50\,Mpc. We find that our constraints for IC\,750 are entirely consistent with the existing constraints on Iron line emission strength and intrinsic luminosity for all SMBH-powered Compton-thick AGN within this volume (see Section~\ref{subsec:ctsamp} and Figure~\ref{fig:ctcomp}).
\end{enumerate}

One Compton-thick IMBH is insufficient to discern whether IC\,750 is the tip of a hitherto unknown population or simply a single remarkable object. However, the general agreement between the existing Compton-thick AGN population and IC\,750 with respect to line-of-sight column density and Eddington ratio (see~Figure~\ref{fig:effedd}), as well as Iron K$\alpha$ equivalent width and intrinsic luminosity (see~Figure~\ref{fig:ctcomp}) suggests that the presence and structure of dense circum-nuclear obscuration may be similar between accreting SMBHs and IMBHs. Dense circum-nuclear obscuration may thus prove to be an additional and necessary consideration in the pursuit of ever-increasing samples of IMBHs in the future.

\section*{ACKNOWLEDGEMENTS}

We thank the anonymous referee for their constructive comments that helped to improve the manuscript. PGB would also like to thank Megan Urry, Colin Burke and Francesca Civano for useful discussions.

The work of DS was carried out at the Jet Propulsion Laboratory, California Institute of Technology, under a contract with NASA.

RJA was supported by FONDECYT grant number 1231718 and by the ANID BASAL project FB210003.

RWP gratefully acknowledges support through an appointment to the NASA Postdoctoral Program at Goddard Space Flight Center, administered by ORAU through a contract with NASA. 

CR acknowledges support from Fondecyt Regular grant 1230345, ANID BASAL project FB210003 and the China-Chile joint research fund.

NTA acknowledges funding from NASA under contracts 80NSSC23K0484 and 80NSSC23K1611.

This research has made use of data obtained from the Chandra Data Archive and the Chandra Source Catalog, and software provided by the Chandra X-ray Center (CXC) in the application packages CIAO and Sherpa.

This research is based on observations made with the NASA/ESA Hubble Space Telescope obtained from the Space Telescope Science Institute, which is operated by the Association of Universities for Research in Astronomy, Inc., under NASA contract NAS 5–26555. These observations are associated with program(s)

This research has made use of the NASA/IPAC Extragalactic Database (NED), which is operated by the Jet Propulsion Laboratory, California Institute of Technology, under contract with the National Aeronautics and Space Administration.

This research has made use of NASA’s Astrophysics Data System Bibliographic Services.

This paper employs a list of Chandra datasets, obtained by the Chandra X-ray Observatory, contained in~\dataset[DOI:10.25574/cdc.287]{https://doi.org/10.25574/cdc.287}.


%

\vspace{5mm}
\facilities{\textit{Chandra}, \textit{Hubble Space Telescope}}


\software{This paper made extensive use of \texttt{matplotlib} \citep{Hunter2007}, \texttt{pandas} \citep{reback2020pandas, mckinney-proc-scipy-2010} and \texttt{astropy} \citep{astropy:2013,astropy:2018,astropy:2022}.
}





\bibliography{bibliography}{}
\bibliographystyle{aasjournal}



\end{document}

%% file: rate_info.tex
\begin{tabular}{ccclccccccc}
\toprule
Obs.\,ID &          PI &                             Obs. start & $\Delta$\,t & T & $\mathcal{C}_{\rm soft}$ & $\mathcal{C}_{\rm hard}$ & $\mathcal{C}_{\rm broad}$ & $\mathcal{S}_{\rm soft}$ & $\mathcal{S}_{\rm hard}$ & $\mathcal{S}_{\rm broad}$ \\
     (1) &     (2) &                                    (3)\,UT &                   (4) &       (5)\,ks &           (6)\,ct\,/\,ks &           (7)\,ct\,/\,ks &            (8)\,ct\,/\,ks &                      (9) &                     (10) &                      (11) \\
\hline
   17006 & J.\,Darling &2014-Oct-05,\,01:16 &                \ldots &          29.7 &        2.75\,$\pm$\,0.31 &        0.82\,$\pm$\,0.17 &         3.61\,$\pm$\,0.35 &                     10.5 &                      5.4 &                      11.8 \\
   22966 &     I.\,Zaw & 2020-Jul-30,\,10:53 &       $+$\,5.8\,years &          48.5 &        1.31\,$\pm$\,0.16 &        0.81\,$\pm$\,0.13 &         2.16\,$\pm$\,0.21 &                      9.1 &                      6.8 &                      11.4 \\
   22967 &     I.\,Zaw & 2021-Feb-22,\,00:56 &      $+$\,6.8\,months &          29.7 &        1.37\,$\pm$\,0.22 &        0.91\,$\pm$\,0.18 &         2.31\,$\pm$\,0.28 &                      7.3 &                      5.7 &                       9.3 \\
   24968 &     I.\,Zaw & 2021-Feb-22,\,16:49 &      $+$\,15.9\,hours &          19.7 &        1.06\,$\pm$\,0.23 &        0.99\,$\pm$\,0.23 &         2.25\,$\pm$\,0.34 &                      5.2 &                      4.8 &                       7.4 \\
   22603 &     I.\,Zaw & 2021-Jul-25,\,23:37 &      $+$\,5.0\,months &          24.8 &        1.08\,$\pm$\,0.21 &        0.90\,$\pm$\,0.19 &         2.02\,$\pm$\,0.29 &                      5.9 &                      5.1 &                       7.9 \\
   25095 &     I.\,Zaw & 2021-Jul-28,\,09:19 &        $+$\,2.4\,days &          24.8 &        1.20\,$\pm$\,0.22 &        0.94\,$\pm$\,0.20 &         2.22\,$\pm$\,0.30 &                      6.2 &                      5.3 &                       8.3 \\
  Co-add &      \ldots &                                 \ldots &                \ldots &         177.1 &        1.50\,$\pm$\,0.09 &        0.87\,$\pm$\,0.07 &         2.43\,$\pm$\,0.12 &                     18.7 &                     13.5 &                      23.3 \\[0.1cm]
\hline
\end{tabular}

%% file: params.tex
{\setlength{\extrarowheight}{5pt}%
\begin{tabular}{rccccl}
\hline
                                          Property &                   \texttt{texrav} &                 \texttt{borus02} &                 \texttt{UXCLUMPY} &               \texttt{warpeddisk} &                    Units \\[5pt]
\hline
                                     log\,$kT^{a}$ &  -0.04$^{+0.10}_{-0.13}$ &  -0.04$^{+0.16}_{-0.72}$ &       -0.04\,$\pm$\,0.10 &  -0.04$^{+0.13}_{-0.19}$ &                      keV \\
                 log\,$L_{0.5-2\,{\rm keV}}$$^{b}$ &  38.05$^{+0.17}_{-0.41}$ &  37.96$^{+0.29}_{-2.15}$ &  38.15$^{+0.05}_{-0.35}$ &  38.00$^{+0.17}_{-0.52}$ &            erg\,s$^{-1}$ \\
                                    $\Gamma$$^{c}$ &   1.85$^{+0.15}_{-0.18}$ &   1.85$^{+0.12}_{-0.22}$ &   1.85$^{+0.16}_{-0.17}$ &   1.85$^{+0.16}_{-0.17}$ &                       -- \\
                      log\,$N_{\rm H,\,los}$$^{d}$ &  24.14$^{+0.74}_{-0.27}$ &  25.02$^{+0.48}_{-0.66}$ &  25.63$^{+0.36}_{-1.36}$ &  24.99$^{+1.01}_{-0.38}$ &                cm$^{-2}$ \\
                         log\,$f_{\rm scat}$$^{e}$ &   0.57$^{+0.43}_{-0.31}$ &  -2.81$^{+2.55}_{-0.36}$ &   0.53$^{+0.34}_{-1.01}$ &  -0.71$^{+0.63}_{-0.77}$ &                       \% \\
            log\,$F_{2-10\,{\rm keV,\,obs}}$$^{f}$ & -13.46$^{+0.08}_{-0.09}$ & -13.47$^{+0.14}_{-0.10}$ & -13.42$^{+0.07}_{-0.09}$ & -13.40$^{+0.05}_{-0.10}$ & erg\,s$^{-1}$\,cm$^{-2}$ \\
            log\,$L_{2-10\,{\rm keV,\,obs}}$$^{g}$ &  38.92$^{+0.08}_{-0.09}$ &  38.91$^{+0.14}_{-0.10}$ &  38.96$^{+0.07}_{-0.09}$ &  38.98$^{+0.05}_{-0.10}$ &            erg\,s$^{-1}$ \\
            log\,$L_{2-10\,{\rm keV,\,int}}$$^{h}$ &  39.87$^{+0.23}_{-0.34}$ &  43.25$^{+0.44}_{-2.43}$ &  40.96$^{+0.43}_{-0.54}$ &  41.55$^{+0.80}_{-0.52}$ &            erg\,s$^{-1}$ \\[5pt]
\hline
        log\,$\mathcal{R}_{2-10\,{\rm keV}}$$^{i}$ &   1.01$^{+0.19}_{-0.39}$ &   4.44$^{+0.44}_{-2.68}$ &   2.04$^{+0.36}_{-0.59}$ &        2.68\,$\pm$\,0.65 &                       -- \\
                          log\,$L_{\rm bol}$$^{j}$ &  41.22$^{+0.43}_{-0.30}$ &  44.63$^{+0.49}_{-2.45}$ &  42.42$^{+0.41}_{-0.62}$ &  43.00$^{+0.83}_{-0.54}$ &            erg\,s$^{-1}$ \\
                    log\,$\lambda_{\rm Edd}$$^{k}$ &  -1.64$^{+0.32}_{-0.44}$ &   1.76$^{+0.50}_{-2.32}$ &  -0.58$^{+0.47}_{-0.56}$ &   0.07$^{+0.80}_{-0.59}$ &                       -- \\[5pt]
\hline
                             $N_{\rm steps}$$^{l}$ &                     1000 &                     2000 &                     1000 &                     1000 &                       -- \\
                        $\overline{\rm RJD}$$^{m}$ &                     1.11 &                     1.13 &                     1.13 &                     1.11 &                       -- \\
                            $f$(RJD)\,$>$\,1$^{n}$ &                   68.5\% &                   69.6\% &                   71.5\% &                   67.7\% &                       -- \\[5pt]
\hline
$\mathcal{Z}$\,/\,$\mathcal{Z_{\tt texrav}}$$^{o}$ &                        1 &      367$^{+247}_{-181}$ &         83$^{+63}_{-43}$ &      468$^{+335}_{-217}$ &                       -- \\[5pt]
\hline
\end{tabular}
}